\documentclass{article}

\usepackage{amsfonts, amsmath, amssymb, amsthm}

\newtheorem{theorem}{Theorem}
\newtheorem{cnj}[theorem]{Conjecture}

\theoremstyle{definition}
\newtheorem{definition}{Definition}

\theoremstyle{remark}
\newtheorem{remark}{Remark}
\newtheorem{example}{Example}

\author{Igor Korepanov}
\title{Deformation of a $3\to 3$ Pachner move relation capturing exotic second homologies}
\date{January 2012}

\begin{document}

\maketitle

\begin{abstract}
A new relation in Grassmann algebra is presented, corresponding naturally to the four-dimensional Pachner move $3\to 3$. This relation is obtained by deforming a known relation associated with an exotic chain complex built for a triangulated four-manifold. The possible deformation terms turn out to correspond to second (middle) homologies of one more exotic complex. Both sides of the new relation are nonlinear in the deformation: they contain terms of degrees 0, 1 and~2.
\end{abstract}

\tableofcontents

\section{Introduction}\label{s:i}

Many results in three-manifold theory have been obtained using Reidemeister torsions, see, for instance, textbook~\cite{t}. The torsions considered in~\cite{t} use abelian representations of the fundamental group, which means that it is the first homologies of a manifold that play essential role. Then, it is stated in~\cite[Theorem~21.4]{t} that this kind of torsion is almost equivalent to the Seiberg--Witten invariant.

This is of course an important equivalence that may give a hint for a right direction of research also in the four-dimensional world. The fact that a $\mathrm{Spin}^c$ structure on a four-manifold is related to its \emph{second} homology group suggests that it may make sense to try and include second homologies in Reidemeister torsions or related objects.

In the present short note, a step is proposed in this direction: some values --- simplex weights --- are constructed, satisfying an algebraic relation imitating a Pachner move --- an elementary rebuilding of a four-manifold triangulation. Specifically, the considered move is $3\to 3$, but there is a hope that these values will satisfy also relations imitating other Pachner moves and, consequently, manifold invariants can be built from them. These values belong to a Grassmann algebra, and algebraic relations between them belong to the Grassmann--Berezin calculus of anti-commuting variables.

First we introduce ``undeformed'' values, linked directly to torsions of an exotic chain complex constructed out of a four-manifold triangulation. Then we ``deform'' them in such way that the deformed values satisfy the same $3\to 3$ algebraic relation, and the possible deformations turn out to be parametrized by homologies in the middle (second) term of another exotic chain complex. Some calculations show that the homologies of our complexes, although being exotic, are closely related with usual homologies.

Theorem proofs in the present note consist usually in direct calculations. These calculations have been really done by the author; it will hopefully be clear from the context when they are done on paper and when with the help of a computer. More conceptual proofs and more details will be presented in further papers.

\section{Exotic chain complex associated with undeformed relations}\label{s:cm}

Let there be a compact oriented triangulated four-manifold~$M$ with boundary. We number all vertices in the triangulation from~$1$ to their total number~$N_0$, and ascribe to each vertex~$i$ its \emph{coordinate}~$\zeta_i$ --- element of some field~$\mathbb F$. These coordinates must be different for different vertices: $\zeta_i\neq \zeta_j$ if $i\neq j$, and we will be using the notation $\zeta_{ij}\stackrel{\rm def}{=}\zeta_i-\zeta_j$.

We are going to define a chain complex of $\mathbb F$-vector spaces and their linear mappings, which we will write as
\begin{equation}\label{cfV}
0 \rightarrow V_0 \stackrel{f_2}{\rightarrow} V_2 \stackrel{f_3}{\rightarrow} V_3 \stackrel{f_4}{\rightarrow} V_4 \stackrel{f_5}{\rightarrow} V_0^* \rightarrow 0
\end{equation}
or, for the reasons that will be clear soon, as
\begin{multline}\label{cf}
0 \rightarrow \begin{pmatrix} \text{inner} \\ \text{vertices} \end{pmatrix} \stackrel{f_2}{\rightarrow} \begin{pmatrix} \text{inner} \\ \text{triangles} \end{pmatrix} \\
\stackrel{f_3}{\rightarrow} \begin{pmatrix} 2\times \\ \text{tetrahedra} \end{pmatrix} \stackrel{f_4}{\rightarrow} \begin{pmatrix} 3\times \\ \text{4-simplices} \end{pmatrix} \stackrel{f_5}{\rightarrow} \begin{pmatrix} \text{inner} \\ \text{vertices} \end{pmatrix} \rightarrow 0 .
\end{multline}

By definition, \emph{inner} vertices of the triangulation form the basis of each of the spaces $V_0$ and~$V_0^*$.

\begin{remark}\label{r:0}
Below in Section~\ref{s:ch2}, it will be convenient for us to use separate notations for vertices as such and basis vectors. In that style, we could say here that vertices are in bijective correspondence with chosen fixed bases in both $V_0$ and~$V_0^*$, and denote the basis vector corresponding to a vertex~$i$ as $\mathrm e_i$ or~$\mathrm e_i^*$, respectively.
\end{remark}

The remaining vector spaces $V_k$, $k=2,3,4$, are defined as follows. First, for each of these~$k$ we introduce vector space~$W_k$ consisting, if $k=3$ or~$4$, of all pairs $(a,i)$, where $a$ is a $k$-simplex and $i\in a$ --- its vertex, and of all such pairs where $a$ is an \emph{inner} triangle if $k=2$. Coordinates of an arbitrary vector $\mathrm y\in W_k$ in this basis are denoted~$y_{a,i}$. Next, we define~$V_k$ as the subspace of~$W_k$ consisting of vectors~$\mathrm y$ whose coordinates obey conditions
\begin{equation}\label{a}
\left.\begin{array}{rcl}
\displaystyle \sum_{i\in a} y_{a,i} &=& 0, \\[3ex]
\displaystyle \sum_{i\in a} \zeta_i y_{a,i} &=& 0
\end{array}\right\} \text{ for every } a .
\end{equation}

Now we pass to the definition of linear mappings $f_2,\dots,f_5$. We will be using letter~$s$ for triangles, $t$ for tetrahedra, and $u$ for 4-simplices. Coordinates of vectors in all spaces will be denoted by the same letter~$y$; this makes no confusion because the subscripts show to which space the vector belongs.

Mapping~$f_2$ makes the vector with following coordinates~$y_{s,i}$ from a vector in~$V_0$ having coordinates~$y_i$:
\begin{equation}\label{f2}
y_{s,i} = (\zeta_{ij}^{-1}-\zeta_{ik}^{-1}) y_i - \zeta_{ij}^{-1} y_j + \zeta_{ik}^{-1} y_k,
\end{equation}
where, first, $i\in s$ (of course), second, $j$ and $k$ are two other vertices of~$s$, and third --- which exactly of these is called $j$ or~$k$ is determined by requiring that the permutation between $ijk$ and the sequence of these vertices ordered according to their numbering must be even.

\begin{remark}\label{r:f2}
A vector obtained according to~\eqref{f2} satisfies of course the conditions~\eqref{a}.
\end{remark}

Mapping~$f_3$ makes the vector with following coordinates~$y_{t,i}$ from a vector in~$V_2$ having coordinates~$y_{s,i}$:
\begin{equation}\label{f3}
y_{t,i} = \sum_{\substack{s\subset t\\ s\ni i}} \epsilon_s^{(t)} y_{s,i},
\end{equation}
where $\epsilon_s^{(t)}=1$ if $s$ enters the boundary of~$t$ with the sign plus, and $\epsilon_s^{(t)}=-1$ otherwise; here the orientations of both $s$ and~$t$ are determined by the order of vertices, and this order is determined by our fixed numbering of vertices.

Mapping~$f_4$ is defined in direct analogy with~$f_3$, by the formula
\begin{equation}\label{f4}
y_{u,i} = \sum_{\substack{s\subset u\\ t\ni i}} \epsilon_t^{(u)} y_{t,i}.
\end{equation}

Finally, mapping~$f_5$ makes the vector with following coordinates~$y_i^*$ from a vector in~$V_4$ having coordinates~$y_{u,i}$:
\begin{equation}\label{f5}
y_i^* = \sum_{u\ni i} \epsilon_u y_{u,i},
\end{equation}
where $\epsilon_u=1$ if the orientation of~$u$ determined by the order of its vertices coincides with one induced by the fixed orientation of the manifold, and $\epsilon_u=-1$ otherwise.

\begin{remark}\label{r:ijklm}
We will also use notations of simplices by their vertices; it is implied, by default, that the vertices are written in the increasing order. So, if there is a 4-simplex $u=ijk\ell m$, then $i<j<k<\ell<m$, and similarly for triangles and tetrahedra.
\end{remark}

\begin{remark}\label{r:m}
Spaces $V_2$, $V_3$ and~$V_4$ have natural isomorphisms with spaces $\tilde V_2$ spanned on the first basis vectors~$(s,i)$ for each inner triangle~$s$ (that is, $i$ is the vertex in~$s$ with the smallest number), $\tilde V_3$ spanned on the first two basis vectors $(t,i)$ and~$(t,j)$ for each tetrahedron~$t$, and~$\tilde V_4$  spanned on the first three basis vectors for each 4-simplex~$u$, respectively. Spaces $V_k$ and~$\tilde V_k$ are identified using the projection along the rest of basis vectors, that is, two last vectors for each $s$, $t$ or~$u$.

Using these projections, basis vectors in $\tilde V_2$, $\tilde V_3$ and~$\tilde V_4$ give basis vectors for $V_2$, $V_3$ and~$V_4$. Thus there are distinguished bases in $V_2$, $V_3$ and~$V_4$, and mappings $f_2,\dots, f_5$ are naturally identified with their matrices.
\end{remark}

\begin{theorem}\label{th:cf}
The sequence \eqref{cfV}, \eqref{cf} of vector spaces and linear mappings is indeed a chain complex, that is,
\[
f_5f_4=0,\qquad f_4f_3=0,\qquad f_3f_2=0.
\]
\end{theorem}

\begin{proof}
Direct calculation.
\end{proof}

\section{Undeformed Grassmann algebraic relation for Pachner move $3\to 3$}\label{s:u}

We consider the Pachner move transforming three adjacent 4-simplices into three other 4-simplices occupying together the same place in the triangulation. We say that the initial three 4-simplices belong to the left-hand side of the move, and the final three --- to its right-hand side. In terms of their vertices, there are 4-simplices $12345$, $12346$ and $12356$ in the l.h.s., and $12456$, $13456$ and $23456$ in the r.h.s.

We now consider the cluster of three 4-simplices in either l.h.s.\ or r.h.s.\ of this move $3\to 3$ as a manifold with boundary, and the chain complex \eqref{cfV}, \eqref{cf} written for it. As this manifold contains no inner vertices, the complex will be shorter by two arrows $f_2$ and~$f_5$. Also, we perform a \emph{gauge transformation} of the remaining arrows $f_3$ and~$f_4$, just to make our formulas consistent with our paper~\cite{ff} and earlier preprint~\cite{pre}. Namely, we multiply each column of matrix~$f_4$ corresponding to each 3-face~$ijk\ell$ by~$\zeta_{k\ell}$ (recall that matrix~$f_4$ has two columns for each 3-face). We denote the resulting matrix~$\tilde f_4$. Accordingly, we divide each row of matrix~$f_3$ by~$\zeta_{k\ell}$, if it corresponds to a 3-face~$ijk\ell$. Additionally, we multiply each column of~$f_3$ by~$\zeta_{jk}$ if it corresponds to an inner 2-face~$ijk$. The obtained matrix is denoted~$\tilde f_3$. This gauge transformation can be also expressed in terms of the obvious scalings of basis vectors in spaces $V_2$ and~$V_3$.

The obtained short complex is
\begin{equation}\label{shc}
0\to V_2 \stackrel{\tilde f_3}{\to} V_3 \stackrel{\tilde f_4}{\to} V_4 \to 0 .
\end{equation}
We would like to exploit the notion of \emph{Reidemeister torsion} for it, but complex~\eqref{shc} is not acyclic. Acyclic are, however, its subcomplexes corresponding to subspaces of~$V_3$ spanned on all basis vectors belonging to inner tetrahedra and the proper number of basis vectors belonging to boundary tetrahedra, so we consider Reidemeister torsions for all such complexes and unite them into one ``generating function'' of Grassmann variables --- \emph{Grassmann weight} of the l.h.s.\ or r.h.s.\ of the Pachner move. The point is that these Grassmann weights for both sides turn out to be the same up to some numeric multiplier, as Theorem~\ref{th:33} below states.

Below in this section we recall the relevant material from paper~\cite{ff}, with some additions and in the form suitable for our aims here. The Grassmann weight of a manifold with boundary can be expressed in terms of weights of individual 4-simplices and Grassmann differential operators corresponding to inner 2-faces. We are going to introduce these objects; they will include also some numeric factors which bring more elegance to the formulas.

Two Grassmann variables $a_{ijk\ell}$ and~$b_{ijk\ell}$ are put in correspondence to each tetrahedron~$ijk\ell$; more exactly, $a_{ijk\ell}$ corresponds to the pair~$(ijk\ell,i)$ of a tetrahedron and its first vertex, and $b_{ijk\ell}$ --- to the pair~$(ijk\ell,j)$ of a tetrahedron and its second vertex. For a given 4-simplex, three linear combinations of these variables are taken, one for each of three rows in matrix~$\tilde f_4$ corresponding to that 4-simplex. The coefficients at $a_{ijk\ell}$ and~$b_{ijk\ell}$ are the corresponding elements of matrix~$\tilde f_4$. For 4-simplex~$12345$, these linear combinations are:
\[
v_{12345,1}=\zeta_{34}a_{1234}-\zeta_{35}a_{1235}+\zeta_{45}a_{1245}-\zeta_{45}a_{1345},
\]
\[
v_{12345,2}=\zeta_{34}b_{1234}-\zeta_{35}b_{1235}+\zeta_{45}b_{1245}+\zeta_{45}a_{2345},
\]
\[
v_{12345,3}=-\zeta_{14}a_{1234}-\zeta_{24}b_{1234}+\zeta_{15}a_{1235}+\zeta_{25}b_{1235}-\zeta_{45}b_{1345}+\zeta_{45}b_{2345}.
\]

In Section~\ref{s:d1}, we will also be using $v_{12345,4}$ and $v_{12345,5}$ determined by the conditions
\begin{equation}\label{v12345}
\left\{\begin{array}{rcl}
v_{12345,1}+v_{12345,2}+v_{12345,3}+v_{12345,4}+v_{12345,5} &=& 0, \\[.75ex]
\zeta_1v_{12345,1} + \zeta_2v_{12345,2} + \zeta_3v_{12345,3} + \zeta_4v_{12345,4} + \zeta_5v_{12345,5} &=& 0.
\end{array}\right.
\end{equation}
The five $v$'s together correspond to the five rows of the following matrix: consider linear operator~$f_4$, written for a manifold consisting of a single tetrahedron~$12345$, as an operator from space~$V_3$ to space~$W_4$ (see the paragraph above formula~\eqref{a}, and recall also Remark~\ref{r:m}), then do for the corresponding $5\times 10$ matrix the same gauge transformation as we did above for the ``usual''~$f_4$.

\begin{remark}\label{r:subs}
Of course we imply that the $v$'s for arbitrary 4-simplex~$ijk\ell m$ are obtained by the same formulas as above, with obvious substitution $1\mapsto i,\allowbreak \dots,\allowbreak 5\mapsto m$. The same applies to similar formulas below.
\end{remark}

\begin{definition}\label{d:W4}
The undeformed Grassmann weight corresponding to 4-sim\-plex $12345$ is the following function of Grassmann variables $a_{i_1i_2i_3i_4}$ and~$b_{i_1i_2i_3i_4}$ attached to each tetrahedron~$i_1i_2i_3i_4$ --- a 3-face of~$12345$:
\begin{equation}\label{W4}
\mathcal W_{12345} \stackrel{\rm def}{=} \frac{1}{\zeta_{45}}v_{12345,1}v_{12345,2}v_{12345,3}.
\end{equation}
\end{definition}

\begin{remark}\label{r:W}
One reason for introducing the factor~$1/\zeta_{45}$ in~\eqref{W4} is that then it follows from~\eqref{v12345} that also
\begin{equation*}
\mathcal W_{12345} = -\frac{1}{\zeta_{35}}v_{12345,1}v_{12345,2}v_{12345,4}=
\dots = \frac{1}{\zeta_{12}}v_{12345,3}v_{12345,4}v_{12345,5}.
\end{equation*}
\end{remark}

As for the matrix~$\tilde f_3$, its elements appear as coefficients at Grassmann partial derivatives in the following definition.

\begin{definition}\label{d:w4}
For a 2-face~$s=ijk$, we introduce the differential operator~$d_{ijk}$ as
\[
d_{ijk} = \sum_{{\rm tetrahedra\ }t\supset s} d_{t,s} ,
\]
where $d_{t,s}$ are the following operators, which we again prefer to write out putting numbers rather than letters in subscripts: we take tetrahedron~$t=1234$ and its four faces, having in mind that, for an arbitrary tetrahedron~$ijkl$ (remember that $i<j<k<l$), the substitution $1\mapsto i,\dots,4\mapsto l$ must be done. So, the operators are:
\begin{equation}\label{2fops4d}
d_{1234,s}= \begin{cases}
(\zeta_{23}/\zeta_{34})\,\partial/\partial a_{1234} - (\zeta_{13}/\zeta_{34})\,\partial/\partial b_{1234} & {\rm if\ } s=123,\\
-(\zeta_{24}/\zeta_{34})\,\partial/\partial a_{1234} + (\zeta_{14}/\zeta_{34})\,\partial/\partial b_{1234} & {\rm if\ } s=124,\\
\partial/\partial a_{1234} & {\rm if\ } s=134,\\
-\partial/\partial b_{1234} & {\rm if\ } s=234.
\end{cases}
\end{equation}
\end{definition}

\begin{theorem}\label{th:33}
The following identity, corresponding naturally to the $3\to 3$ Pachner move, holds:
\begin{multline}\label{33}
\int \mathcal W_{12345} \mathcal W_{12346} \mathcal W_{12356} w_{123} \frac{\mathrm da_{1234}\,\mathrm db_{1234}}{\zeta_{34}} \frac{\mathrm da_{1235}\,\mathrm db_{1235}}{\zeta_{35}}\frac{\mathrm da_{1236}\,\mathrm db_{1236}}{\zeta_{36}} \\
 = \int \mathcal W_{12456} \mathcal W_{13456} \mathcal W_{23456} w_{456} \frac{\mathrm da_{1456} \,\mathrm db_{1456}}{\zeta_{56}} \frac{\mathrm da_{2456} \,\mathrm db_{2456}}{\zeta_{56}} \frac{\mathrm da_{3456} \,\mathrm db_{3456}}{\zeta_{56}} ,
\end{multline}
where
\[
w_{123} = d_{123}^{-1}1, \qquad w_{456} = d_{456}^{-1}1 
\]
are any Grassmann algebra elements of degree~$1$ satisfying differential equations $d_{123}w_{123} = 1$ and $d_{456}w_{456} = 1$, for instance, $w_{123} = \zeta_{23}^{-1}\zeta_{34}a_{1234}$ and $w_{456}=-b_{1456}$.
\end{theorem}

\begin{proof}
Direct calculation.
\end{proof}

\section{Exotic chain complex associated with deformations}\label{s:ch2}

In this section, we present one more exotic complex, built for the same four-manifold triangulation as complex \eqref{cfV}, \eqref{cf}, and using the same vertex coordinates~$\zeta_i$. The homologies in the middle term of this new complex will parametrize the deformations of 4-simplex Grassmann weights~$\mathcal W$, preserving the relation~\eqref{33}, which will be introduced in Section~\ref{s:d1}. The new complex is:
\begin{multline}\label{cg}
0 \rightarrow \begin{pmatrix} \text{inner} \\ \text{vertices} \end{pmatrix} \stackrel{g_2}{\rightarrow} \begin{pmatrix} \text{inner} \\ \text{tetrahedra} \end{pmatrix} \\
\stackrel{g_3}{\rightarrow} \begin{pmatrix} 3\times \\ \text{oriented} \\ \text{4-simplices} \end{pmatrix} \stackrel{g_4}{\rightarrow} \begin{pmatrix} \text{oriented} \\ \text{inner} \\ \text{edges} \end{pmatrix} \stackrel{g_5}{\rightarrow} \begin{pmatrix} 2\times \\ \text{inner} \\ \text{vertices} \end{pmatrix} \rightarrow 0 ,
\end{multline}
and here follows the description of its linear mappings, together with the vector spaces when these may be not obvious.

Below we use different notations for vertices and other simplices, on the one hand, and basis vectors on the other, in accordance with Remark~\ref{r:0}.

Mapping~$g_2$ acts as follows: $\text{vertex }1\mapsto \text{tetrahedron }1234\colon$
\[
g_2(\mathrm e_i)=\sum_{t\ni i} \frac{1}{\zeta_{ij}\zeta_{ik}\zeta_{i\ell}} \mathrm e_t ,
\]
where $\mathrm e_i$ is the basis vector corresponding to an inner vertex~$i$; the sum is taken over all tetrahedra~$t$ containing~$i$; three other vertices of~$t$ are denoted $j$, $k$ and~$\ell$ (with no restriction on the order of numbers $i,\allowbreak j,\allowbreak k,\ell$), and $\mathrm e_t$ is the basis vector corresponding to tetrahedron~$t$.

To explain mapping~$g_3$, we first render more concrete the definition of the middle space ``$3\times{}$oriented 4-simplices''. Recall that an \emph{overfull system} of vectors differs from a basis of a vector space in that although any vector can be presented as a linear combination of vectors from overfull system, this is not done uniquely. The middle space is defined as the linear span of the following overfull system~$\{\mathrm e_{u,i}\}$. Five vectors are assigned to each 4-simplex; these vectors are said to correspond to its five vertices and are denoted, for a simplex~$12345$, as $\mathrm e_{12345,1}$, \dots, $\mathrm e_{12345,5}$. By definition, there are the following, and only the following, linear dependencies between them:
\begin{equation}\label{12345}
\left\{\begin{array}{rcl}
\mathrm e_{12345,1} + \dots + \mathrm e_{12345,5} &=& 0 ,\\[.75ex]
\zeta_1 \mathrm e_{12345,1} + \dots + \zeta_5 \mathrm e_{12345,5} &=& 0 .
\end{array}\right.
\end{equation}

\begin{remark}\label{r:5}
The dependencies~\eqref{12345} are exactly the same as dependencies~\eqref{v12345} between the~$v$'s.
\end{remark}

Mapping $g_3$ acts as follows: let an inner tetrahedron~$1234$ be part of boundaries of 4-simplices $12345$ and $12346$. Then,
\begin{equation}\label{g3}
g_3(\mathrm e_{1234})=\mathrm e_{12345,5}+\mathrm e_{12346,6},
\end{equation}
and it remains to say that, in the case of arbitrary numbers/letters in place of $1\dots 6$, nothing depends on their order or orientation of anything.

\begin{remark}\label{r:g3e}
In Section~\ref{s:d1}, we will need also an analogue of~$g_3$ for \emph{boundary} tetrahedra. This is defined in the obvious way, where there is only one term is the analogue of~\eqref{g3}, as a boundary tetrahedron is part of only one 4-simplex.
\end{remark}

To define~$g_4$, we will need the orientation of each 4-simplex~$ijk\ell m$. To be exact, we introduce numbers $\epsilon_{ijk\ell m}=\pm 1$ the following way: if our default ordering of vertices, namely $i<j<k<\ell<m$, determines an orientation of the 4-simplex coinciding with that induced by the orientation of the manifold, then $\epsilon_{ijk\ell m}=1$, and $\epsilon_{ijk\ell m}=-1$ otherwise. We continue the definition of~$g_4$ on the example of 4-simplex $ijk\ell m=12345$. We also denote this 4-simplex by the single letter~$u$.

If a vector~$\mathrm x$ is a linear combination of $\mathrm e_{u,1}$, \dots, $\mathrm e_{u,5}$:
\begin{equation}\label{x}
\mathrm x = x_{u,1} \mathrm e_{u,1} + \dots + x_{u,5} \mathrm e_{u,5},
\end{equation}
then the coefficients~$x_{u,1},\dots,x_{u,5}$ are not, due to~\eqref{12345}, determined uniquely, but quantities like
\begin{equation}\label{y}
y_{123} = \zeta_{23}x_{u,1} + \zeta_{31}x_{u,2} + \zeta_{12}x_{u,3},
\end{equation}
for an oriented triangle $123\subset 12345$, \emph{are} determined uniquely from~$\mathrm x$. Mapping~$g_4$ acts on the vector~$\mathrm x$ defined by~\eqref{x} as follows:
\begin{equation}\label{g4}
g_4(\mathrm x)=y_{123} \mathrm e_{45} - y_{124} \mathrm e_{35} + \dots + y_{345} \mathrm e_{12} ,
\end{equation}
where $\mathrm e_{ij}$ is the basis vector corresponding to the oriented edge~$ij$, so it changes its sign together with the orientation: $\mathrm e_{ij}=-\mathrm e_{ji}$. The sign at each summand $y_{i_1i_2i_3} \mathrm e_{i_4i_5}$ is the sign of the permutation between $i_1i_2i_3i_4i_5$ and $12345$, and there are of course 10 summands in the r.h.s.\ of~\eqref{g4} --- corresponding to all edges.

Finally, mapping~$g_5$ acts as follows:
\[
g_5(\mathrm e_{12})=\mathrm e_1^* + \zeta_2 \mathrm f_1^* - \mathrm e_2^* - \zeta_1 \mathrm f_2^* ,
\]
where $\mathrm e_i^*$ and $\mathrm f_i^*$ are two basis vectors corresponding to an inner vertex~$i$ in the rightmost space of complex~\eqref{cg}.

\begin{theorem}\label{th:cg}
The sequence \eqref{cg} is indeed a chain complex:
\[
g_5g_4=0,\qquad g_4g_3=0,\qquad g_3g_2=0.
\]
\end{theorem}

\begin{proof}
Direct calculation.
\end{proof}

\section{Deformations}\label{s:d1}

Let a number $x_{u,i}\in \mathbb F$ be given for each pair $(u,i)$, where $u$ is a 4-simplex and~$i\in u$ --- its vertex. We call the set of all such numbers an \emph{$x$-chain}.

\begin{definition}\label{d:d}
The 4-simplex weight deformed by a given $x$-chain is
\begin{equation}\label{tW}
\tilde{\mathcal W}_u = \mathcal W_u + \epsilon_u ( x_{u,i}v_{u,i} + x_{u,j}v_{u,j} + x_{u,k}v_{u,k} + x_{u,\ell}v_{u,\ell} + x_{u,m}v_{u,m} ),
\end{equation}
where $u$ is the 4-simplex with vertices $i<j<k<\ell<m$ (we thus write for brevity $\mathcal W_u=\mathcal W_{ijk\ell m}$ and so on), and $\epsilon_u$ stays for its orientation: $\epsilon_u=1$ if the orientation determined by the order $ijk\ell m$ of vertices agrees with the chosen orientation of the manifold, and $\epsilon_u=-1$ otherwise.
\end{definition}

\begin{example}\label{x:eps33}
For the six 4-simplices taking part in a $3\to 3$ move, $\epsilon_{12345}=1$, $\epsilon_{12346}=-1$, $\epsilon_{12356}=1$, $\epsilon_{12456}=1$, $\epsilon_{13456}=-1$, and $\epsilon_{23456}=1$.
\end{example}

We call an $x$-chain a \emph{cycle} if its image under the mapping~$g_4$ is zero (that is, the zero chain of inner edges), and we call it a \emph{boundary} if it is the image of some inner tetrahedron chain due to mapping~$g_3$.

\begin{example}\label{x:cb33}
For any side of the $3\to 3$ move, any $x$-chain is a cycle, simply because there are no inner edges. Moreover, any $x$-chain in this case can be obtained by the analogue of mapping~$g_3$ from a tetrahedron chain involving \emph{all} --- not only inner --- tetrahedra, compare Remark~\ref{r:g3e}.
\end{example}

\begin{theorem}\label{th:d1}
Consider the relation~\eqref{33} with the $\mathcal W$'s replaced with $\tilde{\mathcal W}$'s defined by~\eqref{tW} using some $x$-chains for its l.h.s.\ and r.h.s. This modified relation~\eqref{33} holds if the mentioned $x$-chains in both sides come from the same \emph{boundary} tetrahedron chain according to the analogue of mapping~$g_3$ introduced in Remark~\ref{r:g3e}.
\end{theorem}

\begin{proof}
Direct calculation.
\end{proof}

\begin{theorem}\label{th:b}
Consider the expression in either l.h.s.\ or r.h.s.\ of the relation~\eqref{33} with the $\mathcal W$'s replaced with $\tilde{\mathcal W}$'s defined by~\eqref{tW} using a given $x$-chain~$\mathcal X$. Then this expression does not change if a boundary $x$-chain is added to~$\mathcal X$.  
\end{theorem}

\begin{proof}
Direct calculation.
\end{proof}

\begin{remark}\label{r:d1}
The undeformed terms in both sides of our algebraic relation corresponding to Pachner move $3\to 3$ are Grassmann algebra elements of degree~$4$. The deformed terms are of degree~$2$ --- these are linear in deformation, that is, in coefficients~$x_{u,i}$ of an $x$-chain, and of degree~$0$ --- these are quadratic in deformation. This means that each of Theorems \ref{th:d1} and~\ref{th:b} states an equality (linear and) \emph{nonlinear} in the deformation.
\end{remark}

\begin{cnj}\label{j:24}
The analogues of Theorems \ref{th:d1} and~\ref{th:b} hold also for a $2\to 4$ Pachner move.
\end{cnj}

\section{Discussion}\label{s:d}

\textbf{Comments about chain complexes. }\vadjust{\nobreak}
The chain complex \eqref{cfV}, \eqref{cf} in actually related to the affine Lie group $\mathrm{Aff}(\mathbb F)$; its Lie algebra $\mathfrak{aff}(\mathbb F)$ will appear in its longer versions, that also include two more linear mappings $f_1$ and~$f_6$ (and this was the reason why the numbering of mappings in \eqref{cfV}, \eqref{cf} starts from~$f_2$).

Somewhat surprisingly, the deformation complex~\eqref{cg} is related, in a similar way, to the larger group $\mathrm{PSL}(2,\mathbb F)$.

\medskip

\textbf{Plans for further research. }\vadjust{\nobreak}
The plans include studying all Pachner moves and doing calculations for specific manifolds, using --- and developing --- our GAP package PL~\cite{PL}.

\end{document}